\documentclass[aps,twocolumn,reprint,superscriptaddress,citeautoscript,longbibliography]{revtex4-2}

\usepackage[dvipdfmx]{graphicx}
\usepackage{amsmath,amssymb}
\usepackage{fixmath}
\usepackage[normalem]{ulem}
\usepackage{bm}
\usepackage{color}
\newcommand{\red}[1]{{#1}}

\usepackage{comment}

\def\lesssim{\ \raise.3ex\hbox{$<$}\kern-0.8em\lower.7ex\hbox{$\sim$}\ }
\def\gesim{\ \raise.3ex\hbox{$>$}\kern-0.8em\lower.7ex\hbox{$\sim$}\ }

\begin{document}

\title{Nonequilibrium noise as a probe of pair-tunneling transport \\ in the BCS--BEC Crossover}

\author{Hiroyuki Tajima}
\email[]{hiroyuki.tajima@phys.s.u-tokyo.ac.jp}
\affiliation{Department of Physics, Graduate School of Science, The University of Tokyo, Tokyo, 113-0033, Japan}
\author{Daigo Oue}
\affiliation{The Blackett Laboratory, Department of Physics, Imperial College London, Prince Consort Road, Kensington, London SW7 2AZ, United Kingdom}
\affiliation{
Kavli Institute for Theoretical Sciences, University of Chinese Academy of Sciences, Beijing, 100190, China.
}
\author{Mamoru Matsuo}
\affiliation{
Kavli Institute for Theoretical Sciences, University of Chinese Academy of Sciences, Beijing, 100190, China.
}
\affiliation{CAS Center for Excellence in Topological Quantum Computation, University of Chinese Academy of Sciences, Beijing 100190, China}
\affiliation{Advanced Science Research Center, Japan Atomic Energy Agency, Tokai, 319-1195, Japan}
\affiliation{RIKEN Center for Emergent Matter Science (CEMS), Wako, Saitama 351-0198, Japan}
\author{Takeo Kato}
\affiliation{The Institute for Solid State Physics, The University of Tokyo, Kashiwa 277-8581, Japan}

\date{\today}

\begin{abstract}
The detection of elementary carriers in transport phenomena is one of the most important keys to understand non-trivial properties of strongly-correlated quantum matter.
Here we propose a method to identify the tunneling current carrier in strongly interacting fermions from nonequilibrium noise in the Bardeen-Cooper-Schrieffer to Bose--Einstein condensate crossover. The noise-to-current ratio, the Fano factor, can be a crucial probe for the current carrier. Bringing strongly-correlated fermions into contact with a dilute reservoir produces a tunneling current in between. The associated Fano factor increases from one to two as the interaction becomes stronger, reflecting the fact that the dominant conduction channel changes from the quasiparticle tunneling to the pair tunneling.
\end{abstract}

\maketitle

Transport phenomena have contributed to the development of the fundamental physics in previous centuries.
Various unconventional phenomena such as superfluidity and superconductivity were observed using transport measurements.
However, clarifying the microscopic mechanism of the transport phenomena in strongly-correlated systems remains challenging because of their complexities such as strong interactions, lattice geometries, as well as multiple degrees of freedom.

\red{Recently, an ultracold atomic system has been regarded as a quantum simulator for strongly-correlated many-body systems such as unconventional superconductors and nuclear systems,
owing to its controllability of physical parameters (e.g., interparticle interactions and lattice structures) and its cleanness~\cite{RevModPhys.80.885,RevModPhys.82.1225}.}
In particular, state-of-the-art experiments for tunneling current have been conducted in strongly interacting Fermi gases~\cite{Krinner8144,hausler2017scanning,kwon2020strongly,luick2020ideal,PhysRevLett.126.055301,PhysRevX.11.021034}.
\red{Moreover, thermoelectric transport has been demonstrated experimentally in an ultracold Fermi gas~\cite{brantut2013thermoelectric}.
A quantum point contact has also been implemented for atomic superfluid junctions~\cite{husmann2015connecting}.}
These experiments motivate us to study tunneling transport associated with the Josephson effect and Cooper-pair tunneling in the superfluid phase of the Bardeen-Cooper-Schrieffer (BCS) to Bose--Einstein-condensate (BEC) crossover~\cite{PhysRevA.77.043609,PhysRevLett.117.255302,PhysRevA.98.041601,damanet2019reservoir,PhysRevA.100.063601,piselli2020josephson,PhysRevResearch.2.023340,setiawan2021analytic,PhysRevResearch.4.023231}.
\red{Such a direction are recently referred to as {\it atomtronics}~\cite{amico2021roadmap}.}

One crucial problem is to understand how strong correlations affect the conduction mechanism, which is necessary for future development of quantum-transport technology. 
Recently, several theoretical efforts have been paid to understand an anomalous tunneling current induced by pairing fluctuations in the normal phase~\cite{PhysRevLett.118.105303,PhysRevA.95.013623,PhysRevResearch.2.023152,furutani2020strong}, as observed in  experiments~\cite{Krinner8144,hausler2017scanning,kwon2020strongly,luick2020ideal,PhysRevLett.126.055301,PhysRevX.11.021034}.
It is anticipated that such anomalous pair-tunneling currents can be induced by the nonlinear tunneling processes~\cite{PhysRevLett.118.105303}, tunneling of a closed-channel molecule~\cite{PhysRevA.95.013623},
and the proximity effect associated with two-body interactions~\cite{PhysRevA.106.033310}.
However, regardless of these different origins, the existence of the pair-tunneling current itself is still an important pending problem because it is difficult to distinguish quasiparticle- and pair-tunneling currents experimentally.
%\red{(pair-tunnelingを調べることの重要性を述べる)}
In this sense, it is worth exploring clear evidence for
anomalous pair currents in a strongly interacting Fermi gas.
%above the superfluid critical temperature.
%\red{(熱量あげてpair-tunnelingをdetectする方法を提案)}

\begin{figure}
    \centering
    \includegraphics[width=6.5cm]{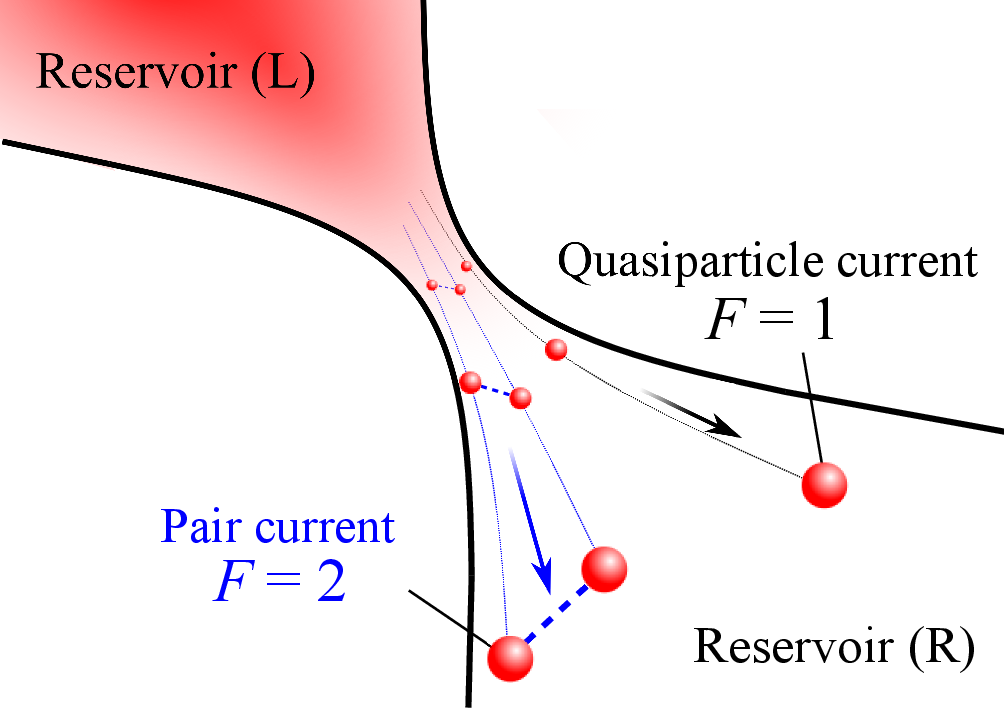}
    \caption{Strongly interacting quantum gases (reservoirs ${\rm L}$ and ${\rm R}$) with a large chemical-potential bias in between.
    The Fano factor $F$ can be regarded as an indicator of the current carrier, i.e., quasiparticle current ($F=1$) and the pair current ($F=2$).
    }
    \label{fig:1}
\end{figure}

For this purpose, measuring the Fano factor is promising, which is defined by a current and the associated nonequilibrium noise~\cite{blanter2000shotnoise,martin2005shotnoise}. The Fano factor in the large-biased setup reflects the effective charge per elementary transport process regardless of system's detail.
The most fascinating example is the detection of fractional charges in fractional quantum Hall systems~\cite{picciotto1997FQHE,Saminadayar1997FQHE}. The Fano factor has been used to determine the effective charge (or spin) in various physical systems such as superconductors~\cite{Jehl1997SC,PhysRevLett.84.3398}, Kondo quantum dots~\cite{PhysRevB.77.241303,Ferrier2016QD}, and magnetic junctions~\cite{PhysRevLett.116.146601,PhysRevB.94.014419,PhysRevLett.120.037201,PhysRevB.97.014427}.
Once the Fano factor is measured in strongly interacting Fermi gases, the existence of the pair-tunneling current will be revealed in an unbiased way.

In this study, we show that the Fano factor $F$ can be used as a probe for the current carrier in the BCS--BEC crossover.
Figure~\ref{fig:1} shows a schematic setup of the large-biased system.
Using the many-body $T$-matrix approach (TMA)~\cite{STRINATI20181,OHASHI2020103739},
we numerically calculate the current and nonequilibrium noise within the Schwinger--Keldysh approach in the two-terminal tunneling junction under a large bias.
We reveal how the Fano factor $F$ changes in a strongly-interacting regime, thereby reflecting the change of the dominant carrier.
In particular, the change of $F$ is a crucial evidence for the pair-tunneling current.
Our result can be tested by cold-atom experiments for which the noise measurement has been theoretically proposed~\cite{PhysRevA.98.063619}.
Moreover, the Fano factor provides direct information of pair-fluctuation effects rather than other measurements such as spin susceptibility and photoemission spectra previously studied in this field~\cite{mueller2017review}.
The current-noise measurement can also be used to identify the carriers of the BCS--BEC crossover in condensed-matter systems such as FeSe semimetal~\cite{lubashevsky2012shallow,kasahara2014field,rinott2017tuning,hanaguri2019quantum}, lithium-intercalated layered nitrides~\cite{PhysRevB.98.064512,nakagawa2021gate}, magic-angle twisted trilayer graphene~\cite{park2021tunable}, and organic superconductor~\cite{PhysRevX.12.011016}.
Moreover, the noise measurement has recently been conducted in a copper oxide heterostructure~\cite{zhou2019electron,bovzovic2020pre} and disordered superconductor~\cite{doi:10.1126/science.abe3987}. 

In the following, we take $\hbar=k_{\rm B}=1$ and consider a unit volume.

%\section*{Guide to using this template on Overleaf}

%Please note that whilst this template provides a preview of the typeset manuscript for submission, to help in this preparation, it will not necessarily be the final publication layout. For more detailed information please see the \href{https://www.pnas.org/page/authors/format}{PNAS Information for Authors}.

%If you have a question while using this template on Overleaf, please use the help menu (``?'') on the top bar to search for \href{https://www.overleaf.com/help}{help and tutorials}. You can also \href{https://www.overleaf.com/contact}{contact the Overleaf support team} at any time with specific questions about your manuscript or feedback on the template.

\section*{Tunneling current and noise}
We consider the Hamiltonian $H=H_{\rm L}+H_{\rm R}+H_{\rm 1T}+H_{\rm 2T}$.
The reservoir Hamiltonian $H_{{\rm j}={\rm L,R}}$ is given by
\begin{align}
    H_{\rm j}=\sum_{\bm{p},\sigma}\xi_{\bm{p},{\rm j}}c_{\bm{p},\sigma,{\rm j}}^\dag c_{\bm{p},\sigma,{\rm j}}+g\sum_{\bm{q}}P_{\bm{q},{\rm j}}^\dag P_{\bm{q},{\rm j}},
    \label{eq:H_i}
\end{align}
where $\xi_{\bm{p},{\rm j}}=p^2/(2m)-\mu_{\rm j}$ denotes the kinetic energy measured from the chemical potential $\mu_{\rm j}$ and $c_{\bm{p},\sigma,{\rm j}}$ denotes the annihilation operator of a Fermi atom with momentum $\bm{p}$ and the pseudospin $\sigma=\uparrow,\downarrow$. 
The second term in Eq.~(\ref{eq:H_i}) denotes the attractive interaction with a contact-type coupling $g$, where $P_{\bm{q},{\rm j}}=\sum_{\bm{p}}c_{-\bm{p}+\bm{q}/2,\downarrow,{\rm j}}c_{\bm{p}+\bm{q}/2,\uparrow,{\rm j}}$ is the pair-annihilation operator and
$g$ is related to the scattering length $a$ as
$\frac{m}{4\pi a}=\frac{1}{g}+\sum_{\bm{p}}\frac{m}{p^2}$~\cite{OHASHI2020103739}.

The one-body tunneling Hamiltonian,
\begin{align}
    H_{\rm 1T}=\sum_{\bm{p},\bm{k},\sigma}\left[t_{\bm{p},\bm{k}}c_{\bm{p},\sigma,{\rm L}}^\dag c_{\bm{k},\sigma,{\rm R}}+{\rm h.c.}\right],
\end{align}
is associated with the one-body potential barrier, where $t_{\bm{p},\bm{k}}$ denotes its coupling strength.
The two-body tunneling Hamiltonian reads
\begin{align}
    H_{\rm 2T}=\sum_{\bm{q},\bm{q}'}\left[w_{\bm{q},\bm{q}'}P_{\bm{q},{\rm L}}^\dag P_{\bm{q}',{\rm R}} + {\rm h.c.} \right],
\end{align}
where $w_{\bm{q},\bm{q}'}$ is the two-body coupling strength, induced by the local interaction term in Eq.~(\ref{eq:H_i}) combined with the one-body potential barrier ~\cite{PhysRevA.106.033310}.
%\textcolor{red}{In the two-channel model for the narrow Feshbach resonance, it can be associated with the tunneling of a closed-channel molecule~\cite{PhysRevA.95.013623}.}
Such two-body tunneling processes can also be obtained within
the multiple one-body tunneling processes in the non-linear regime~\cite{PhysRevLett.118.105303,PhysRevA.100.043604,PhysRevResearch.2.023340,furutani2020strong}.
We note that regardless of their origins, these two-body tunnelings induce the pair-tunneling current.
Similar tunneling effects have also been examined in one-dimensional few-body systems~\cite{Sowinski_2016,PhysRevA.98.053614}.
Here, we do not go into details on the origin of the one- and two-body tunneling, but rather investigate their possible consequence in observable quantities.
\red{However, we emphasize that the two-body tunneling term is necessary to describe the molecule tunneling in the deep BEC side (and therefore the entire crossover), where the pair-tunneling induced by the higher-order one-body tunneling process is suppressed due to the reduced dissociation of molecules with the large binding energy~\cite{furutani2020strong}. In Fig.~S1 of the supplement~\cite{Supplement}, we estimate the tunneling couplings in the case of delta-function-like potential barrier~\cite{griffiths2018introduction,PhysRevResearch.4.023231} based on Ref.~\cite{PhysRevA.106.033310}.}

Using the Schwinger--Keldysh approach, we evaluate the expectation values of the current operator $\hat{I}=i[\hat{N}_{\rm L},H]$ ($\hat{N}_{\rm j}=\sum_{\bm{p},\sigma}c_{\bm{p},\sigma,{\rm j}}^\dag c_{\bm{p},\sigma,{\rm j}}$ denotes the density operator in the ${\rm j}$-reservoir) in the steady state at the lowest-order tunneling couplings by a sum of the one- and two-body contributions as $I=I_{\rm qp}+I_{\rm pair}$, where each component reads~\cite{PhysRevA.106.033310,Supplement}
\begin{align}
\label{eq:I}
    {I_{\rm qp}} &=\int_{-\infty}^{\infty} \frac{d\omega}{2\pi} \sum_{\bm{p},\bm{k},\sigma} |t_{\bm{k},\bm{p}}|^2 \mathcal{A}_{\bm{k},{\rm L}}(\omega) \mathcal{A}_{\bm{p},{\rm R}}(\omega)\cr
&\qquad\qquad \times [f_{\rm L}(\omega)-f_{\rm R}(\omega)], \cr
    {I_{\rm pair}} &= 2\int_{-\infty}^{\infty}   \frac{d\omega}{2\pi}\sum_{\bm{q},\bm{q}'} |w_{\bm{q},\bm{q}'}|^2 \mathcal{B}_{\bm{q},{\rm L}}(\omega) \mathcal{B}_{\bm{q}',{\rm R}}(\omega)\cr
&\qquad\qquad \times [b_{\rm L}(\omega)-b_{\rm R}(\omega)]. 
\end{align}
In Eq.~(\ref{eq:I}), {$\mathcal{A}_{\bm{k},{\rm j}}(\omega)$ and $\mathcal{B}_{\bm{q},{\rm j}}(\omega)$ denote one- and two-particle spectral functions, respectively,} $f_{\rm j}(\omega)$ and $b_{\rm j}(\omega)$ denotes the Fermi and Bose distribution functions, and $\mu_{{\rm b},{\rm j}}=2\mu_{\rm j}$ denotes the bosonic-pair chemical potential in the ${\rm j}$-reservoir.
For the detection of the pair-tunneling current, it is crucial to consider the small tunneling coupling regime where the nonequilibirum noise reflects an effective particle number in tunneling process.~\footnote{We note that the validity of the truncation with respect to the lowest-order tunneling coupling was confirmed in the recent experiment~\cite{PhysRevLett.126.055301}.
}

\red{We define the current noise as
    $\bar{\mathcal{S}}(t_1,t_2)=\frac{1}{2}\langle \hat{I}(t_1)\hat{I}(t_2)+\hat{I}(t_2)\hat{I}(t_1)\rangle$~\cite{PhysRevB.46.12485,blanter2000shot,imry2002introduction,bouchiat2005nanophysics} [see also, e.g., Ref.~\cite{PhysRevLett.120.037201}].
For the steady-state transport with the time-translational symmetry, we assume that the noise depends on $t_1-t_2$ as $\bar{\mathcal{S}}(t_1,t_2)\equiv \bar{\mathcal{S}}(t_1-t_2)$ (being independent of $\frac{t_1+t_2}{2}$). 
Its Fourier component reads 
\begin{align}
\label{eq:add1}
    \bar{\mathcal{S}}(\omega)&=\frac{1}{\tau}\int_0^{\tau}dt_1\int_0^{\tau}dt_2e^{i\omega(t_1-t_2)}\bar{\mathcal{S}}(t_1-t_2),
\end{align}
where $\tau$ is the typical time scale for the noise measurement.
Taking $t=t_1-t_2$ and $\bar{\mathcal{S}}(t)=\frac{1}{2}\langle \hat{I}(t)\hat{I}(0)+\hat{I}(0)\hat{I}(t)\rangle$, we obtain the zero-frequency limit of the noise power $\mathcal{S}\equiv \bar{\mathcal{S}}(\omega\rightarrow \eta)$ ($\eta$ is an infinitesimally small number) as}
\begin{align}
    \mathcal{S}=\frac{1}{2}\int_{-\infty}^{\infty}
    dt\left(\langle\hat{I}(t)\hat{I}(0)\rangle+\langle\hat{I}(0)\hat{I}(t)\rangle\right),
\end{align}
\red{where we considered the limit of $\tau\rightarrow\infty$. In this regard, we briefly note that $\tau$ should be sufficiently longer than the transport timescale $\tau_0$, where in the recent experiment $\tau_0= O(10^{-1})$ {\rm s} is found~\cite{brantut2013thermoelectric}.}
Similar to the calculation above, we can evaluate the current noise~\cite{Supplement} as the sum of the two contributions:
$\mathcal{S} = \mathcal{S}_{\rm qp} + \mathcal{S}_{\rm pair}$,
where
\begin{align}
\label{eq:S}
    {\mathcal{S}_{\rm qp}} &=\int_{-\infty}^{\infty}
    \frac{d\omega}{2\pi}\sum_{\bm{p},\bm{k},\sigma}
    |t_{\bm{k},\bm{p}}|^2\mathcal{A}_{\bm{k},{\rm L}}(\omega)
    \mathcal{A}_{\bm{p},{\rm R}}(\omega)\cr
    &\quad\times\left[f_{{\rm L}}(\omega)\{1-f_{{\rm R}}(\omega)\}+\{1-f_{{\rm L}}(\omega)\}f_{{\rm R}}(\omega)\right],
       \cr
    {\mathcal{S}_{\rm pair}} &=4\int_{-\infty}^{\infty}
    \frac{d\omega}{2\pi}
    \sum_{\bm{q},\bm{q}'}|w_{\bm{q},\bm{q}'}|^2\mathcal{B}_{\bm{q},{\rm L}}(\omega)
    \mathcal{B}_{\bm{q}',{\rm R}}(\omega)\cr
    &\quad\times
    \left[b_{\rm L}(\omega)\{1+b_{\rm R}(\omega)\}
    +b_{\rm R}(\omega)\{1+b_{\rm L}(\omega)\}
    \right].
\end{align}
\red{The bias between the reservoirs is included in the distribution function and therefore \eqref{eq:S} is valid for the case with the temperature bias~\cite{lumbroso2018electronic}.}
In the large \red{chemical potential} bias limit ($\Delta\mu\equiv \mu_{\rm L}-\mu_{\rm R}\rightarrow \infty$), we can prove $\mathcal{S}_{\rm qp}/I_{\rm qp}=1$ and $\mathcal{S}_{\rm pair}/I_{\rm pair}=2$ without any further approximations~\cite{Supplement}.
This motivates us to consider the Fano factor,
\begin{align}
    F=\frac{\mathcal{S}}{I} = {\frac{\mathcal{S}_{\rm qp}+\mathcal{S}_{\rm pair}}{I_{\rm qp}+I_{\rm pair}}}.
\end{align}
The Fano factor $F$ changes from 1 to 2, according to whether the quasiparticle or pair tunneling is dominant and hence, it is a useful probe for the current carrier.
In particular, the Fano factor $F$ becomes 1 and 2 in the BCS limit ($a^{-1}\rightarrow -\infty$) and BEC limit ($a^{-1}\rightarrow \infty$), respectively.
Importantly, the deviation of $F$ from 1 indicates a clear evidence of the pair-tunneling process yet to be not well understood in cold atomic systems~\cite{PhysRevA.106.033310}.
Therefore, the observation of $F$ can be a crucial key for understanding transport phenomena in strongly interacting systems.

In this study, we consider 
the large bias regime (see Fig.~\ref{fig:1}) characterized by $\mu_{\rm L}-\mu_{\rm R}\rightarrow \infty$~\cite{Supplement,PhysRevA.91.013606}
and the momentum-conserved tunneling processes as $t_{\bm{p},\bm{k}}=\mathcal{T}_1\delta_{\bm{p},\bm{k}}$
and $w_{\bm{q},\bm{q}'}=\mathcal{T}_2\delta_{\bm{q},\bm{q}'}$, for simplicity.
To see the qualitative behavior of $F$, we use the spectral functions $\mathcal{A}_{\bm{k},{\rm j}}(\omega)=-2\, {\rm Im}\, G_{\bm{k},{\rm j}}(i\omega_n\rightarrow\omega-\mu_{\rm j}+i\eta)$ and $\mathcal{B}_{\bm{q},{\rm j}}(\omega)=-2\, {\rm Im}\, \mathcal{G}_{\bm{q},{\rm j}}(i\nu_\ell\rightarrow \omega-\mu_{{\rm b},{\rm j}}+i\eta)$ with an infinitesimal small number $\eta$,
where thermal single- and two-particle propagators $G_{\bm{k},{\rm j}}(i\omega_n)$ and $\mathcal{G}_{\bm{q},{\rm j}}(i\nu_\ell)$ with fermion and boson Matsubara frequencies $i\omega_n$ and $i\nu_\ell$ are evaluated within the many-body TMA~\cite{zwerger2011bcs,PhysRevB.61.15370} (see also Supplemental Material~\cite{Supplement}).
We employ $\eta=10^{-2}E_{\rm F,L}$ in the numerical calculation to avoid the divergent behavior of the current associated with the momentum-conserved tunneling in the weak- and strong-coupling limits,
where $E_{\rm F,L}=(3\pi^2 N_{\rm L})^{\frac{2}{3}}/(2m)$ denotes the Fermi energy of the ${\rm L}$ reservoir with the number density $N_{\rm L}$.
However, our result can be qualitatively unchanged by this treatment because the distribution functions play a key role in determining $F$ rather than the detailed structures of tunneling junctions.
Moreover, $\mathcal{T}_2$ must be normalized to suppress the ultraviolet divergence in $B_{\bm{q},{\rm j}}(\omega)$.
For this purpose, we introduce the renormalized two-body tunneling coupling $\mathcal{T}_{2,{\rm ren.}}=\frac{\Lambda^2k_{\rm F,{\rm L}}}{3\sqrt{2}\pi^2}\mathcal{T}_2$ where $k_{\rm F,L}=\sqrt{2mE_{\rm F,L}}$ denotes the Fermi momentum.
Such a divergence can also be avoided by introducing the form factor for the relative momentum $\bm{p}$ in $P_{\bm{q},{\rm j}}$~\cite{PhysRevB.68.144507}. 
In this work, we take $\Lambda=100k_{\rm F,L}$~\cite{OHASHI2020103739} in the practical calculation.
This value is associated with the effective range $r_{\rm eff}$ as $r_{\rm eff}=\frac{4}{\pi\Lambda}$~\cite{OHASHI2020103739}.

\section*{Fano factor thoughout the BCS-BEC crossover}
\begin{figure}[t]
    \centering
    \includegraphics[width=0.8\linewidth
]{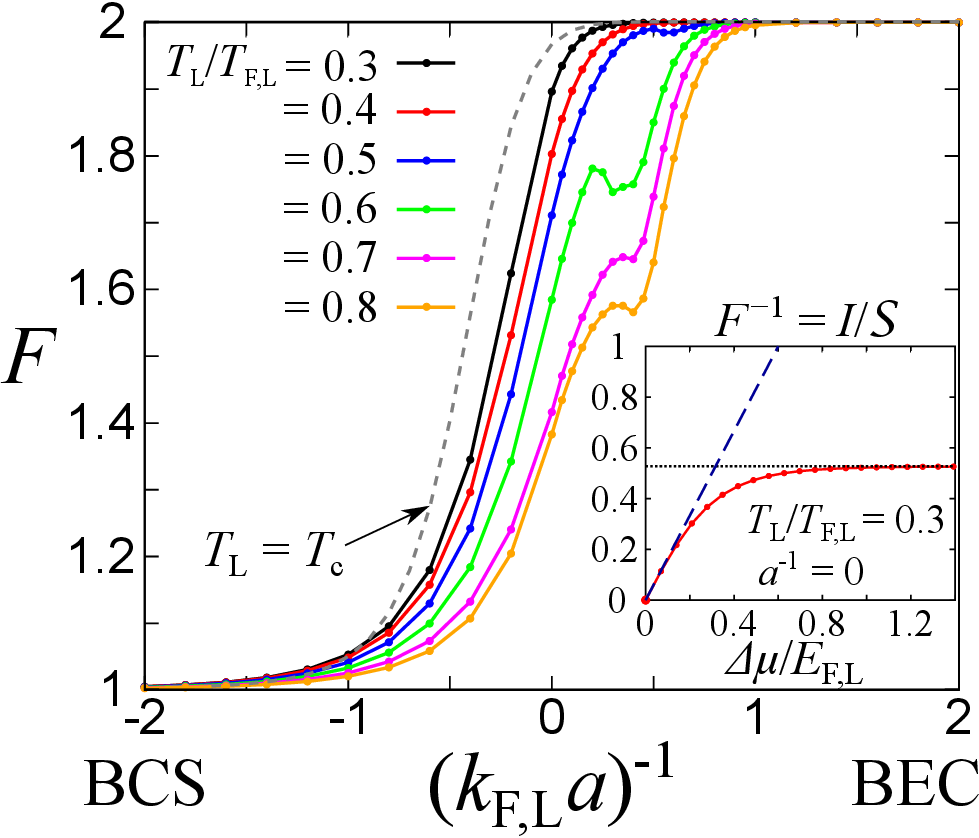}
    \caption{Fano factor $F$, associated with tunneling transport between two reservoirs, throughout the BCS-BEC crossover for various temperatures $T_{\rm L}$ in the reservoir ${\rm L}$. 
    The reservoir ${\rm R}$ is almost vacuum.
    The ratio between tunneling couplings is given as $\mathcal{T}_{2,{\rm ren.}}/\mathcal{T}_1=1$.
    For comparison, we plot the result at $T_{\rm L}=T_{\rm c}$ (dashed curve).
    Note that $T_{\rm c}$ changes in the range of $0.02T_{\rm F,L}\lesssim T_{\rm c} \lesssim 0.24T_{\rm F,L}$ depending on $(k_{\rm F,L}a)^{-1}$.
    The inset shows the bias ($\Delta\mu$) dependence of $F^{-1}$ at $T_{\rm L}/T_{\rm F,L}=0.3$ and $a^{-1}=0$.
    The dashed and dotted lines represent the Onsager's relation $F^{-1}(\Delta\mu\rightarrow 0)=\frac{\Delta\mu}{2T}$~\cite{Supplement} and the large bias limit, respectively.
    }
    \label{fig:2}
\end{figure}
Fig.~\ref{fig:2} shows the Fano factor $F$ as a function of the dimensionless interaction parameter $(k_{\rm F,L}a)^{-1}$ in the entire BCS-BEC crossover regime above the superfluid critical temperature $T_{\rm c}$.
We considered $\mathcal{T}_{2,{\rm ren.}}/\mathcal{T}_1=1$, and the reservoir ${\rm R}$ was regarded as almost vacuum ($\mu_{\rm L}-\mu_{\rm R}\rightarrow \infty$)~\cite{Supplement}.
As we showed in the inset of Fig.~\ref{fig:2}, the large-bias assumption can be justified when $\Delta\mu$ is larger than a typical many-body scale in the reservoir (i.e., $E_{\rm F,L}$).
One can clearly see that $F$ evolves from 1 to 2 with increasing the interaction strength in Fig.~\ref{fig:2}, indicating that the current carrier gradually changes from quasiparticles ($F=1$) to pairs ($F=2$).
Such a behavior is universal in the sense that these asymptotic values do not depend on any details on the model parameters and  structures of tunneling junctions.
More explicitly, at the large bias limit, one can obtain~\cite{Supplement}
\begin{align}
\label{eq:F-largebias}
F(\Delta\mu\rightarrow \infty)\rightarrow \frac{I_{\rm qp}+2I_{\rm pair}}{I_{\rm qp}+I_{\rm pair}},
\end{align}
where $I_{\rm qp}$ and $I_{\rm pair}$ denote the contributions of the quasiparticle and pair tunnelings, respectively. 
The Fano factor $F$ approaches 1 and 2 in the quasiparticle-dominant ($I_{\rm qp}\gg I_{\rm pair}$) and pair-dominant regimes ($I_{\rm pair}\gg I_{\rm qp}$), respectively.
Although the interaction dependence of the Fano factor $F$ is deeply related to properties of the tunneling junctions and spectral functions of the carriers, one can find from Eq.~(\ref{eq:F-largebias}) that $F\rightarrow 1$ ($F\rightarrow 2$) in the limit of $a^{-1}\rightarrow -\infty$ ($a^{-1}\rightarrow \infty$) regardless of the detailed properties of the system. 
Moreover, $F=2$ can be realized even above $T_{\rm c}$ because of strong interactions leading to the formation of preformed Cooper pairs in the BCS--BEC crossover. 
With increasing the temperature, $F$ tends to be suppressed because thermal effects assist the dissociation of pairs. 
Nevertheless, even at finite temperature, $F$ approaches 2 with increasing the interaction because bound molecules are dominant in the deep BEC regime~\footnote{\red{Here, ``BEC regime" is used for the regime where the two-body attraction is so strong that the associated superfluid state behaves like molecular BEC below $T_{\rm c}$~\cite{STRINATI20181,OHASHI2020103739}. In this regard, the strongly-attractive regime even above $T_{\rm c}$ is also referred as to the BEC regime for characterizing the interaction strength.} }
where $T_{\rm L}\lesssim E_{\rm b}$
[$E_{\rm b}=1/(ma^2)$ is the two-body binding energy].

\begin{figure}[t]
    \centering
    \includegraphics[width=0.8\linewidth
]{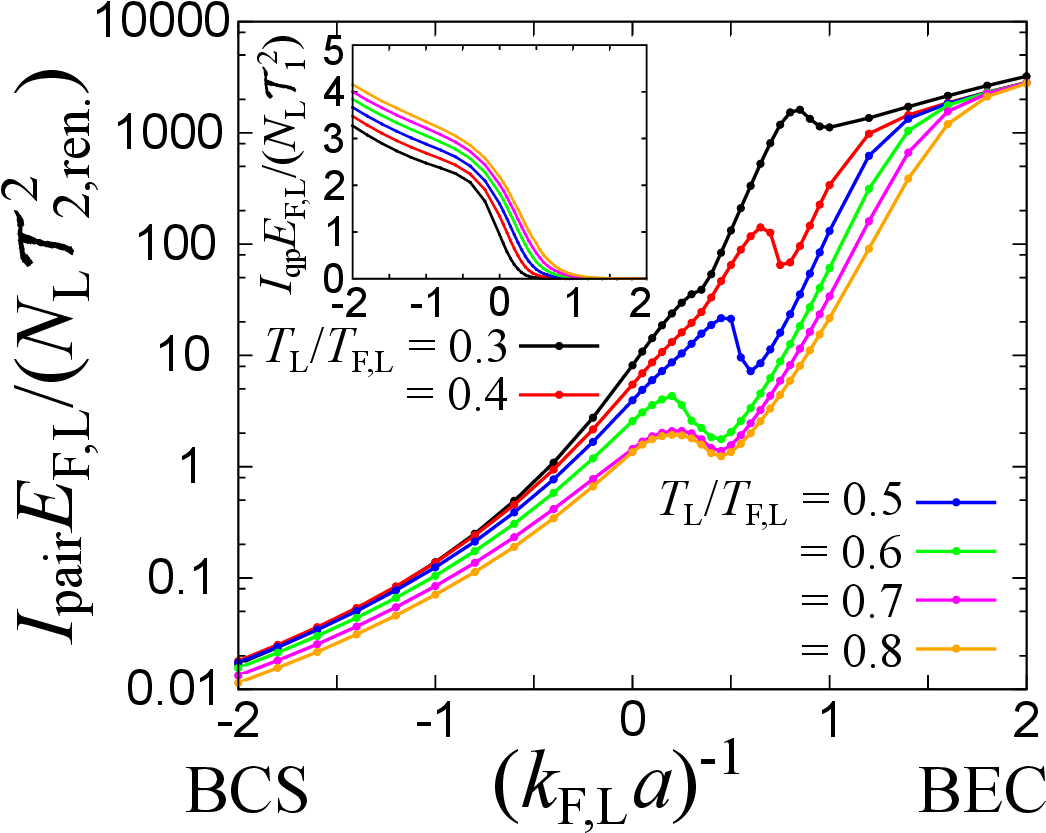}
    \caption{Pair-tunneling current $I_{\rm pair}$ in the normal phase throughout the BCS-BEC crossover at different temperatures. The inset shows the quasiparticle current $I_{\rm qp}$ with the same horizontal axis $(k_{\rm F,L}a)^{-1}$. }
    \label{fig:3}
\end{figure}

To see 
the detailed behavior of the Fano factor $F$, 
we plot $I_{\rm qp}$ and $I_{\rm pair}$ throughout the BCS-BEC crossover at different temperatures in Fig.~\ref{fig:3}.
From the inset of Fig.~\ref{fig:3}, the quasiparticle current $I_{\rm qp}$ is exponentially suppressed with increasing the attractive interaction.
This suppression (in particular, the rapid drop of $I_{\rm qp}$ at $(k_{\rm F,L}a)^{-1}\gesim-0.5$) is induced by the pairing fluctuation effect~\cite{OHASHI2020103739}, i.e., the reduction of $\mathcal{A}_{\bm{k},{\rm L}}(\omega)$ near $|\bm{k}|=k_{\rm F,L}$ and $\omega=E_{\rm F,L} \ (\simeq \mu_{\rm L})$ by the particle-hole coupling.
We note that this fluctuation effects result in the pseudogap in the density of state near $T_{\rm c}$~\cite{mueller2017review}.
Finally, $I_{\rm qp}$ approaches zero in the BEC limit ($(k_{{\rm F},{\rm L}} a)^{-1}\rightarrow \infty$) because of the formation of molecules with large binding energies.
These results are qualitatively consistent with previous work~\cite{PhysRevLett.118.105303,furutani2020strong}.
On the other hand, $I_{\rm pair}$ drastically increases with increasing the interaction strength $(k_{{\rm F},{\rm L}} a)^{-1}$ as shown in Fig.~\ref{fig:3}.
At the BCS side ($(k_{{\rm F},{\rm L}} a)^{-1}< 0$)
where the attraction is not strong to form a two-body bound state in vacuum,
the contribution of $I_{\rm pair}$ can be regarded as the tunneling of the preformed Cooper pairs into the two-body continuum in the reservoir ${\rm R}$.
In the strong-coupling BEC regime ($(k_{\rm F,L}a)^{-1} > 1$ and $T_{\rm L}/E_{\rm b}\lesssim 1$),
$I_{\rm pair}$ describes the tunneling transport of bound molecules across two reservoirs, because the two-body bound state exists in the reservoir ${\rm R}$ with the same coupling $g$. 
Such a tunneling current associated with weakly-interacting molecular bosons becomes large due to their long lifetime and the Bose enhancement of low-energy distributions.

One can also see a dip-hump structure of $I_{\rm pair}$ in the intermediate regime.
Here, $\mu_{\rm L}$ is close to zero and changes its sign, indicating that the dominant contribution changes from the preformed-pair transfer to the molecule-to-molecule transport across the junction.
From the unitary limit ($(k_{\rm F,L}a)^{-1}=0$), the preformed-pair transfer increases due to the overlap with the bound-state spectra in $B_{\bm{q},{\rm R}}(\omega)$ and eventually decreases because of the decrease in $\mu_{\rm L}$.
With increasing the interaction further, the inter-reservoir molecule-to-molecule transition emerges where the bound-state spectra in two reservoirs get close to each other in the energy axis $\omega$
~\footnote{We note that in this regime the numerical cost is large due to the overlap of Bose distribution function and sharp peaks in $\mathcal{B}_{\bm{q},{\rm L,R}}$. We confirmed that the qualitative behavior is robust against the accuracy of the frequency integration.}. 
Although these structures reflect the physical properties of the system, they also depend on the detailed setup of the tunneling junctions (e.g., the ratio between the tunneling couplings $\mathcal{T}_{2,{\rm ren.}}/\mathcal{T}_1$)~\cite{Supplement}.

\begin{figure}[t]
    \centering
    \includegraphics[width=0.8\linewidth]{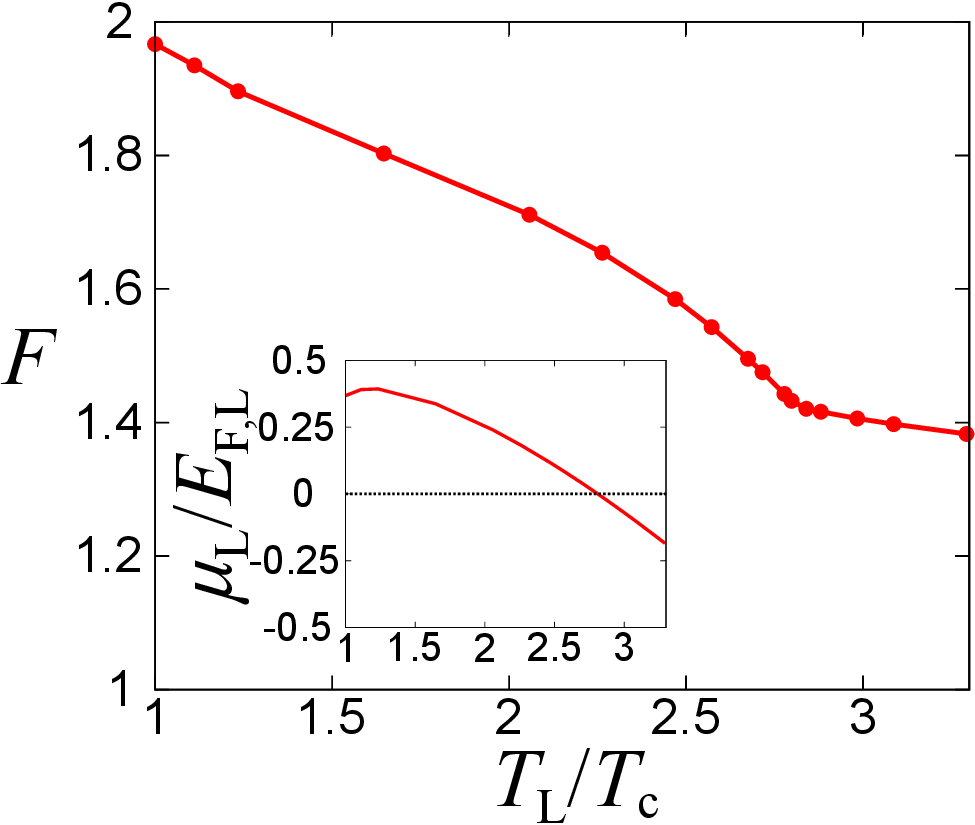}
    \caption{Temperature dependence of the Fano factor $F$ in the unitary limit [$1/(k_{\rm F,L}a)=0$] with $\mathcal{T}_{2,{\rm ren.}}/\mathcal{T}_1=1$.
    The horizontal axis is taken as $T_{\rm L}/T_{\rm c}$, where $T_{\rm c}$ is the superfluid critical temperature.
    The inset shows the chemical potential $\mu_{\rm L}$ as a function of $T_{\rm L}/T_{\rm c}$ for a given Fermi energy $E_{\rm F,L}$.
    }
    \label{fig:4}
\end{figure}

Figure~\ref{fig:4} shows the temperature dependence of {the Fano factor} $F$ in the unitary limit ($(k_{\rm F,L}a)^{-1}=0$).
Because $\mathcal{B}_{\bm{q},{\rm R}}(\omega)$ does not involve a bound molecule pole, the transfer of the preformed Cooper pairs in the reservoir ${\rm L}$ to the two-body continuum in the reservoir ${\rm R}$ can be anticipated in the unitary limit.
One can see the enhancement of the Fano factor $F$ at the low-temperature regime.
In particular, the curvature of the Fano factor $F$ is modified at $T_{\rm L}/T_{\rm c}\simeq 2.8$, where the sign of $\mu_{\rm L}$ changes from negative to positive one as the temperature decreases (see the inset of Fig.~\ref{fig:4}).
Although the Fano factor depends on $\mathcal{T}_{2,{\rm ren.}}/\mathcal{T}_1$ as shown in Fig.~S2 of the supplement~\cite{Supplement}, the qualitative behavior, i.e., suppression of the pair-tunneling current due to increase of the temperature is unchanged regardless of the value of $\mathcal{T}_{2,{\rm ren.}}/\mathcal{T}_1$.
For estimating the value of $\mathcal{T}_{2,{\rm ren.}}/\mathcal{T}_1$ (which depends on the potential barrier and the interaction strength) in each experimental setup, see Ref.~\cite{PhysRevA.106.033310}.
\red{In the supplement~\cite{Supplement}, we show that $\mathcal{T}_{2,{\rm ren.}}/\mathcal{T}_1$ can be tuned and it is possible to realize $\mathcal{T}_{2,{\rm ren.}}/\mathcal{T}_1\simeq 1$ by adjusting the strength of the potential barrier as
$\mathcal{T}_{2,{\rm ren.}}/\mathcal{T}_1\propto
\left(1+\frac{V_0}{E_{\rm F,L}}\right)^{-1}\left[1+
    \left(\frac{V_0}{E_{\rm F,L}}\right)^2(k_{\rm F,L}\ell)^2\right]^{-1/2}$ for the potential barrier given by $V=V_0\delta(x/\ell)$ perpendicular to the $x$ axis ($V_0$ and $\ell$ are the strength and the characteritic length scale of the barrier) }.
%\red{Although the absolute value of $F$ depends on $\mathcal{T}_{2,{\rm ren.}}/\mathcal{T}_1$~\cite{Supplement}, one can see how the pair current becomes dominant from the temperature dependence of $F$ regardless of $\mathcal{T}_{2,{\rm ren.}}/\mathcal{T}_1$.}
At a positive $\mu_{\rm L}$, the pole of the preformed Cooper pairs gradually appears in $\mathcal{B}_{\bm{q},{\rm L}}(\omega)$. 
Thus, the behavior of the Fano factor $F$ can be regarded as a signature of the preformed Cooper pairs.
Because the preformed Cooper pairs play an important role in the pseudogap physics of ultracold Fermi gases~\cite{mueller2017review},
the Fano factor contributes to the further understanding of pairing pseudogaps in the BCS--BEC crossover regime.
Incidentally, because TMA does not capture the self-energy shift in $\Pi_{\bm{q},{\rm L}}(\omega)$,
the curvature change of the Fano factor $F$ may differ from the temperature where $\mu_{\rm L}=0$ in actual experiments and in more sophisticated theoretical approaches~\cite{STRINATI20181,OHASHI2020103739}.
To evaluate the spectral functions, the analytic continuation should be carefully performed in Monte Carlo simulations~\cite{JARRELL1996133}.
We note that because TMA reproduces the second-order virial expansion~\cite{LIU201337},
our result in the relatively high-temperature regime can give an accurate estimate of $F$ for given tunnel couplings.

\section*{Summary} 
In this study, we showed that the Fano factor (i.e., the noise-to-current ratio $F=\mathcal{S}/I$) can be a useful probe for current carriers in the BCS--BEC crossover at large-biased tunneling junctions.
Using the many-body TMA, we demonstrated that the Fano factor $F$ gradually changes from one to two as the interaction strength increases in the normal phase,
indicating that the dominant current carrier changes from the quasiparticle ($F=1$) to the pair ($F=2$) along the BCS-BEC crossover.
Our prediction can be tested by experiments and uncover nonequilibrium strong-coupling physics via transport measurements.
\red{While we have focused on the large bias limit, such a situation can be achieved when the bias is larger than the many-body energy scale (i.e., Fermi energy of the dense reservoir).}
Furthermore, our result indicates that the noise measurement is useful for the study of the BCS-BEC crossover and pair-fluctuation effects in unconventional superconductors.

\begin{acknowledgments}
This work is supported in part by
Grants-in-Aid for Scientific Research from JSPS (Grants Nos.~JP18H05406, JP20K03831, 22K13981).
D.O.~is supported by the President's PhD Scholarships at Imperial College London, by JSPS Overseas Research Fellowship, by the Institution of Engineering and Technology (IET), and by Funda\c{c}\~ao para a Ci\^encia e a Tecnologia and Instituto de Telecomunica\c{c}\~oes under project UIDB/50008/2020.
MM is partially supported by the Priority Program of the Chinese Academy of Sciences, Grant No.~XDB28000000.
This manuscript was posted on a preprint: https://doi.org/10.48550/arXiv.2202.03873
\end{acknowledgments}

\bibliography{./reference}

%supplemental 
\begin{widetext}

\section*{Pair-tunneling coupling}
Following Ref.~\cite{PhysRevA.106.033310}, we obtain the renormalized pair-tunneling coupling as
\begin{align}
    \mathcal{T}_{2,{\rm ren.}}\equiv \frac{\Lambda^2k_{\rm F,L}}{3\sqrt{2}\pi^2}\mathcal{T}_2\simeq
    \frac{\Lambda^2k_{\rm F,L}}{3\sqrt{2}\pi^2}
    2|g| {\rm Re}[B_{0,\uparrow}B_{0,\downarrow}]
\end{align}
where
$B_{0,\sigma}$ is the amplitude of the transmitted wave with respect to the potential barrier.
Since we consider the spin-balanced system, we take $B_{0,\uparrow}=B_{0,\downarrow}\equiv B_0$.
Near unitarity ($a^{-1}\simeq 0$),
$g$ can be rewritten as
    $g=\frac{4\pi a}{m}\frac{1}{1-\frac{4\pi a}{m}\frac{m\Lambda}{2\pi^2}}\simeq -\frac{2\pi^2}{m\Lambda}$.
While $g$ is negative, $\mathcal{T}_{2,{\rm ren.}}$ can be taken to be positive by the appropriate gauge transformation.
In the case of the delta potential barrier $V(x)=V_0\delta(x/\ell)$ being perpendicular to the $x$ axis ($\ell$ is the typical length scale of the tunneling region, e.g., the width of the actual potential barrier), which is given by a constant $V(\bm{k})=V_0$ in the momentum space,
we find~\cite{griffiths2018introduction}
\begin{align}
    {\rm Re}[B_{0}^2]\simeq T_{\rm trans.}=\frac{1}{1+\frac{mV_0^2\ell^2}{2E_{\rm F,L}}},
\end{align}
where $T_{\rm trans.}$ is the transmission coefficient.
For simplicity, we take $E_{\rm F,L}$ for the energy of the incident particle and the transverse motion along the barrier is neglected.
Combining them, we get
\begin{align}
    \mathcal{T}_{2,{\rm ren.}}&\simeq 
    \frac{4\Lambda k_{\rm F,L}}{3\sqrt{2}m}
    \frac{1}{1+\frac{mV_0^2\ell^2}{2E_{\rm F,L}}}.
    %\equiv
    %\frac{16 k_{\rm F,L}}{3\sqrt{2}\pi mr_{\rm eff}}
    %\frac{1}{1+\frac{mV_0^2\ell^2}{2E_{\rm F,L}}}
\end{align}

\red{In turn, we obtain the quasiparticle-tunneling coupling $\mathcal{T}_1$ as
\begin{align}
    \mathcal{T}_1\simeq B_{0}(E_{\rm F,L}+V_0)\equiv\frac{E_{\rm F,L}+V_0}{\sqrt{1+\frac{mV_0^2\ell^2}{2E_{\rm F,L}}}},
\end{align}
where we ignore the higher order term involving the reflection amplitude.
Note that the Hartree term $gN_{\rm L}$ is negligible compared to the other term for the present short-range interaction.
In this way, we obtain
\begin{align}
\label{eq:t2t1ratio}
    \frac{\mathcal{T}_{2,{\rm ren.}}}{\mathcal{T}_1}&\simeq \frac{4}{3\sqrt{2}}\frac{\Lambda k_{\rm F,L}}{m(E_{\rm F,L}+V_0)}
    \frac{1}{\sqrt{1+\frac{mV_0^2\ell^2}{2E_{\rm F,L}}}}\equiv \frac{8}{3\sqrt{2}}\frac{\Lambda/k_{\rm F,L}}{\left(1+\frac{V_0}{E_{\rm F,L}}\right)\sqrt{1+
    \left(\frac{V_0}{E_{\rm F,L}}\right)^2(k_{\rm F,L}\ell)^2}}.
\end{align}
While we use the contact-type interaction with the cutoff regularization,
the cutoff $\Lambda$ can be associated with the effective range $r_{\rm eff}$ as $r_{\rm eff}=\frac{4}{\pi \Lambda}$ (and moreover the interaction range $r_{\rm int.}$)~\cite{OHASHI2020103739}.
In cold atom experiments, the typical interaction range is approximately given by $|k_{\rm F,L}r_{\rm int.}|\simeq 10^{-2}$~\cite{RevModPhys.82.1225,PhysRevLett.109.220402}.
%(e.g.,
%PRL 109, 220402 (2012) and 
%C. Chin RMP). 
}

\begin{figure}[t]
    \centering
    \includegraphics[width=8cm]{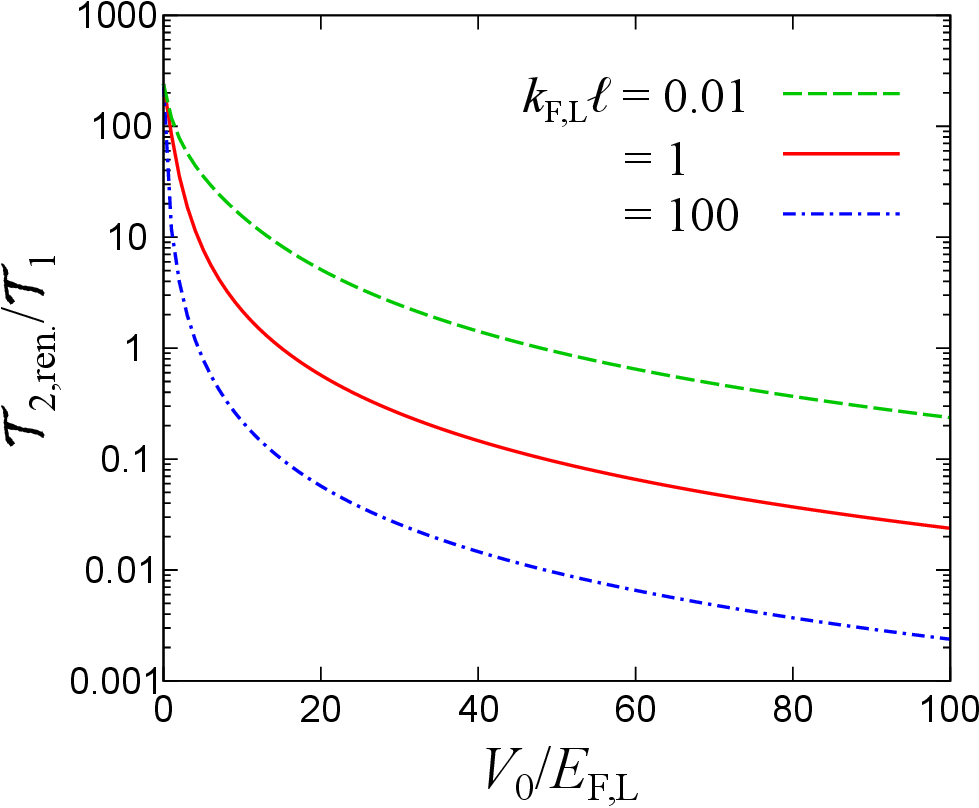}
    \caption{The dimensionless ratio between $\mathcal{T}_{\rm 2, ren.}$ and $\mathcal{T}_1$ given by \eqref{eq:t2t1ratio}, where we used $k_{\rm F,L}r_{\rm eff}=10^{-2}$. $V_0$ and $\ell$ are defined through the delta-function form of the potential barrier $V(x)=V_0\delta(x/\ell)$.}
    \label{fig:add1}
\end{figure}

\red{Figure~\ref{fig:add1} shows $\mathcal{T}_{2,{\rm ren.}}/\mathcal{T}_1$ in Eq.~\eqref{eq:t2t1ratio} as a function of $V_0/E_{\rm F,L}$ at different $k_{\rm F,L}\ell$. One can see that the ratio can be tuned by changing $V_0$.}

\section*{Schwinger-Keldysh approach for current and noise}

We start from the current operator given by
\begin{align}
\label{eq:app_I}
    \hat{I}&=\hat{I}_{\rm qp} + \hat{I}_{\rm pair}, \\
    \hat{I}_{\rm qp}&= i\sum_{\bm{p},\bm{k},\sigma}
    t_{\bm{k},\bm{p}}\left[c_{\bm{k},\sigma,{\rm L}}^\dag c_{\bm{p},\sigma,{\rm R}}-c_{\bm{p},\sigma,{\rm R}}^\dag c_{\bm{k},\sigma,{\rm L}}\right], \\
    \hat{I}_{\rm pair} &= 2i\sum_{\bm{q},\bm{q}'}w_{\bm{q},\bm{q}'}\left[P_{\bm{q},{\rm L}}^\dag P_{\bm{q}',{\rm R}}-P_{\bm{q}',{\rm R}}^\dag P_{\bm{q},{\rm L}}\right],
\end{align}
where $\hat{I}_{\rm qp}$ and $\hat{I}_{\rm pair}$ are operators for quasiparticle and pair currents, respectively.
Truncating the higher-order contributions with respect to the tunneling Hamiltonians [i.e.,$O(H_{\rm 1T}^3)$, $O(H_{\rm 2T}^3)$],
we can evaluate their expectation values, $I_{\rm qp}(t_1,t_2)=\langle\Psi(t_1)| \hat{I}_{\rm qp}|\Psi(t_2)\rangle$ and
$I_{\rm pair}(t_1,t_2)=\langle\Psi(t_1)| \hat{I}_{\rm pair}|\Psi(t_2)\rangle$,
for the different times $t_1$ and $t_2$, where $|\Psi(t)\rangle$ is the state-vector of the steady state.
First, the quasiparticle contribution reads
\begin{align}
\label{eq:app_Iqp}
    I_{\rm qp}(t_1,t_2)
    &=-2\int_Cdt'\sum_{\bm{p},\bm{k},\sigma}
    |t_{\bm{k},\bm{p}}|^2
    \, {\rm Re}\left[
    \langle T_C 
    c_{\bm{k},\sigma,{\rm R}}(t_2)
    c_{\bm{k},\sigma,{\rm R}}^\dag(t')
    \rangle
    \langle T_C c_{\bm{p},\sigma,{\rm L}}(t')c_{\bm{p},\sigma,{\rm L}}^\dag (t_1)\rangle
    \right],
\end{align}
where $C$ denotes the Keldysh contour.
Note that while the right hand side of Eq.~(\ref{eq:app_Iqp}) depends only on $t_1-t_2$ in considering the steady state.
Using the Green's functions,
we rewrite $I_{\rm qp}(t_1,t_2)$ as
\begin{align}
I_{\rm qp}(t_1,t_2)
    =2\int_{-\infty}^{\infty}dt'
    \sum_{\bm{p},\bm{k},\sigma}
    |t_{\bm{k},\bm{p}}|^2
    &\,{\rm Re}\left[
    G_{\bm{p},{\rm R}}^{\rm ret.}(t_2-t')
    G_{\bm{k},{\rm L}}^{<}(t'-t_1)+
        G_{\bm{p},{\rm R}}^{<}(t_2-t')
    G_{\bm{k},{\rm L}}^{\rm adv.}(t'-t_1)
    \right],
\end{align}
where $G^{\rm ret.(adv.)}$ is the retarded (advanced) Green's function of a fermion in thermal equilibrium.
The lesser component $G^{<}$ contains the information of the thermal distribution in each reservoir.
Here, we take $t_1=t_2\equiv t$ and the Fourier transformation
\begin{align}
    I_{\rm qp}
    =2\int \frac{d\omega}{2\pi}
    \sum_{\bm{p},\bm{k},\sigma}
    |t_{\bm{k},\bm{p}}|^2
&{\rm Re}\left[G_{\bm{p},{\rm R}}^{\rm ret.}(\omega)G_{\bm{k},{\rm L}}^{<}(\omega)
+G_{\bm{p},{\rm R}}^{<}(\omega)
G_{\bm{k},{\rm L}}^{\rm ret.*}(\omega)
\right].
\end{align}
Moreover, we use
\begin{align}
\label{eq:app_g1lesser}
    G_{\bm{k},{\rm j}}^{<}(\omega)=-2if_{\rm j}(\omega)\,{\rm Im}\,G_{\bm{k},{\rm j}}^{\rm ret.}(\omega)
    \equiv if_{\rm j}(\omega) \mathcal{A}_{\bm{k},{\rm j}}(\omega)
    , 
\end{align}
where 
\begin{align}
    f_{\rm j}(\omega)=\frac{1}{\exp(\frac{\omega-\mu_{\rm j}}{T_{\rm j}})+1}
\end{align}
is the Fermi-Dirac distribution function.
We use Matsubara Green's functions in each reservoir reaching thermal equilibrium as a grand-canonical ensemble with $-\mu_{\rm j}\hat{N}_{\rm j}$ and obtain the retarded(advanced) Green's function by the analytic continuation with $\mu_{\rm j}$ as
$i\omega_n\rightarrow\omega+i\eta-\mu_{\rm j}$ in each reservoir.
Then, we obtain
\begin{align}
\label{eq:app_I_qp}
    I_{\rm qp}
    &=\int \frac{d\omega}{2\pi} \sum_{\bm{p},\bm{k},\sigma}
    |t_{\bm{k},\bm{p}}|^2
    \mathcal{A}_{\bm{p},{\rm L}}(\omega)
        \mathcal{A}_{\bm{k},{\rm R}}(\omega)
        \left[f_{\rm L}(\omega)-f_{\rm R}(\omega)\right].
\end{align}
Similarly, we obtain the pair current contribution as
\begin{align}
\label{eq:app_I_pair}
    I_{\rm pair}
    =2\sum_{\bm{q},\bm{q}'}\int\frac{d\omega}{2\pi}|\omega_{\bm{q},\bm{q}'}|^2
    \mathcal{B}_{\bm{q},{\rm L}}(\omega)
    \mathcal{B}_{\bm{q}',{\rm R}}(\omega)
    \left[b_{\rm L}(\omega)-b_{\rm R}(\omega)
    \right],
\end{align}
where we used the relation for the two-particle Green's function given $\mathcal{G}^{<}$ by
\begin{align}
\label{eq:app_g2lesser}
    \mathcal{G}_{\bm{q},{\rm j}}^{<}(\omega)=2ib_{\rm j}(\omega)\,{\rm Im}\,\mathcal{G}_{\bm{q},{\rm j}}^{\rm ret.}(\omega)\equiv -ib_{\rm j}(\omega)\mathcal{B}_{\bm{q},{\rm j}}(\omega),
\end{align}
and the Bose-Einstein distribution function
\begin{align}
    b_{\rm j}(\omega)=\frac{1}{\exp\left(\frac{\omega-\mu_{{\rm b},{\rm j}}}{T_{\rm j}}\right)-1},
\end{align}
with a bosonic (pair) chemical potential $\mu_{{\rm b},{\rm j}}=2\mu_{\rm j}$.
$\mathcal{G}^{<(>)}$ and $\mathcal{G}^{\rm ret.(adv.)}$ 
are the lesser (greater) and retarded (advanced) components of two-particle Green's functions, respectively.
One can find that $I=I_{\rm qp}+I_{\rm pair}$ obtained from Eqs.~(\ref{eq:app_I_qp}) and (\ref{eq:app_I_pair}) is equivalent to Eq.~(4) in the main text.
\red{We beriefly note that one may find the correlation between quasiparticle and tunneling currents in the higher-order contributions such as the term proportional to $t_{\bm{k},\bm{p}}^2w_{\bm{q},\bm{q}'}$, which is beyond the scope in this work.}

Next, we consider the current noise
\begin{align}
    \mathcal{S}=\frac{1}{2}\int_{-\infty}^{\infty}
    dt\left(\langle\hat{I}(t)\hat{I}(0)\rangle+\langle\hat{I}(0)\hat{I}(t)\rangle\right).
\end{align}
At lowest order of tunneling couplings, we obtain
\begin{align}
    \langle \hat{I}(t)\hat{I}(0)\rangle
    &=\sum_{\bm{p},\bm{k},\sigma}|t_{\bm{p},\bm{k}}|^2\left[G_{\bm{k},{\rm L}}^{<}(t)G_{\bm{p},{\rm R}}^{>}(-t)+G_{\bm{p},{\rm R}}^{<}(t)G_{\bm{k},{\rm L}}^{>}(-t)\right]\cr
    &-4\sum_{\bm{q},\bm{q}'}|w_{\bm{q},\bm{q}'}|^2
    \left[\mathcal{G}_{\bm{q},{\rm L}}^{<}(t)\mathcal{G}_{\bm{q}',{\rm R}}^{>}(-t)+\mathcal{G}_{\bm{q}',{\rm R}}^{<}(t)\mathcal{G}_{\bm{q},{\rm L}}^{>}(-t)\right],
\end{align}
\begin{align}
    \langle \hat{I}(0)\hat{I}(t)\rangle
    &=\sum_{\bm{p},\bm{k},\sigma}|t_{\bm{p},\bm{k}}|^2\left[G_{\bm{k},{\rm L}}^{<}(-t)G_{\bm{p},{\rm R}}^{>}(t)+G_{\bm{p},{\rm R}}^{<}(-t)G_{\bm{k},{\rm L}}^{>}(t)\right]\cr
    &-4\sum_{\bm{q},\bm{q}'}|w_{\bm{q},\bm{q}'}|^2
    \left[\mathcal{G}_{\bm{q},{\rm L}}^{<}(-t)\mathcal{G}_{\bm{q}',{\rm R}}^{>}(t)+\mathcal{G}_{\bm{q}',{\rm R}}^{<}(-t)\mathcal{G}_{\bm{q},{\rm L}}^{>}(t)\right].
\end{align}
Collecting them and taking the Fourier transformation, we obtain
\begin{align}
    \mathcal{S}&=\mathcal{S}_{\rm qp} + \mathcal{S}_{\rm pair}, \\
    \mathcal{S}_{\rm qp} &=\int_{-\infty}^{\infty}\frac{d\omega}{2\pi}
    \sum_{\bm{k},\bm{p},\sigma}
    |t_{\bm{k},\bm{p},\sigma}|^2
    \left[G_{\bm{k},{\rm L}}^{<}(\omega)G_{\bm{p},{\rm R}}^{>}(\omega)
    +G_{\bm{k},{\rm L}}^{>}(\omega)G_{\bm{p},{\rm R}}^{<}(\omega)
    \right], \cr
    \mathcal{S}_{\rm pair}&=-4\int_{-\infty}^{\infty}\frac{d\omega}{2\pi}
    \sum_{\bm{q},\bm{q}'}|w_{\bm{q},\bm{q}'}|^2\left[\mathcal{G}_{\bm{q},{\rm L}}^{<}(\omega)
    \mathcal{G}_{\bm{q}',{\rm R}}^{>}(\omega)
    +\mathcal{G}_{\bm{q},{\rm L}}^{>}(\omega)
    \mathcal{G}_{\bm{q}',{\rm R}}^{<}(\omega)
    \right].
\end{align}
Using the relations associated with greater Green's functions
\begin{align}
    G_{\bm{p},{\rm j}}^{>}(\omega)=-i\mathcal{A}_{\bm{p},{\rm j}}(\omega)[1-f_{\rm j}(\omega)], \quad
    \mathcal{G}_{\bm{q},{\rm j}}^{>}(\omega)=-i\mathcal{B}_{\bm{q},{\rm j}}(\omega)
    [1+b_{\rm j}(\omega)],
\end{align}
and the lesser ones given by Eqs.~(\ref{eq:app_g1lesser}) and (\ref{eq:app_g2lesser}),
we obtain
\begin{align}
\label{eq:xm}
    \mathcal{S}_{\rm qp} &=\int_{-\infty}^{\infty}
    \frac{d\omega}{2\pi}\sum_{\bm{k},\bm{p},\sigma}
    |t_{\bm{k},\bm{p},\sigma}|^2\mathcal{A}_{\bm{k},{\rm L}}(\omega)
    \mathcal{A}_{\bm{p},{\rm R}}(\omega) \left[f_{{\rm L}}(\omega)\{1-f_{{\rm R}}(\omega)\}+\{1-f_{{\rm L}}(\omega)\}f_{{\rm R}}(\omega)\right]\cr
    \mathcal{S}_{\rm pair}&=4\int_{-\infty}^{\infty}
    \frac{d\omega}{2\pi}
    \sum_{\bm{q},\bm{q}'}
    |w_{\bm{q},\bm{q}'}|^2
    \mathcal{B}_{\bm{q},{\rm L}}(\omega)
    \mathcal{B}_{\bm{q}',{\rm R}}(\omega)
    \left[b_{\rm L}(\omega)\{1+b_{\rm R}(\omega)\}
    +b_{\rm R}(\omega)\{1+b_{\rm L}(\omega)\}
    \right],
\end{align}
which is equivalent to Eq.~(6) in the main text.
\red{We note that, in \eqref{eq:xm},
the terms proportional to $\left[f_{{\rm  j}}(\omega)\{1-f_{{\rm j}}(\omega)\}\right]$ and 
$\left[b_{\rm j}(\omega)\{1+b_{\rm j}(\omega)\}\right]$ (${\rm j}={\rm L,R}$) do not appear in contrast to Ref.~\cite{blanter2000shot} because we consider the lowest-order contributions $O(t_{\bm{k},\bm{p},\sigma}^2)$ and $O(w_{\bm{q},\bm{q}'}^2)$ without the reflection term.
Moreover, the correlation of two noises may appear in the higher-order contributions [e.g., $O(t_{\bm{k},\bm{p},\sigma}^2w_{\bm{q},\bm{q}'})$], which will be considered in the future work.
}
For a small bias limit at equal temperatures $T_{\rm L}=T_{\rm R}\equiv T$ where $\Delta\mu\rightarrow 0$ and $f_{\rm R}(\omega)\rightarrow f_{\rm L}(\omega)\equiv f(\omega)$ with $\mu_{\rm R}\rightarrow \mu_{\rm L}\equiv\mu$, we obtain
\begin{align}
    f_{\rm L}(\omega)-f_{\rm R}(\omega)
    &=-\frac{\partial f(\omega)}{\partial \omega}\Delta\mu+O((\Delta\mu)^2),
\end{align}
\begin{align}
    b_{\rm L}(\omega)-b_{\rm R}(\omega)
        &=-2\frac{\partial b(\omega)}{\partial \omega}\Delta\mu+O((\Delta\mu)^2).
\end{align}
Using 
\begin{align}
    f(\omega)\{1-f(\omega)\}
    =-T\frac{\partial f(\omega)}{\partial \omega},
\quad
    b(\omega)\{1+b(\omega)\}
    =-T\frac{\partial b(\omega)}{\partial \omega},
\end{align}
we recover the Onsager's relation
\begin{align}
\label{eq:Onsager}
    \mathcal{S}(\Delta\mu\rightarrow 0)=2T\frac{I}{\Delta\mu}.
\end{align}

Moreover, the current and the noise can be rewritten as
\begin{align}
\label{eq:app_I2}
    I_{\rm qp}
    &=\int_{-\infty}^{\infty}
    \frac{d\omega}{2\pi}
    \sum_{\bm{p},\bm{k},\sigma}
    |t_{\bm{k},\bm{p}}|^2
    \mathcal{A}_{\bm{k},{\rm L}}(\omega)
        \mathcal{A}_{\bm{p},{\rm R}}(\omega)
        \left[-\frac{1}{2}\frac{\sinh\left(\frac{\beta_{\rm L}(\omega-\mu_{\rm L})-\beta_{\rm R}(\omega-\mu_{\rm R})}{2}\right)}{\cosh\left(\frac{\beta_{\rm L}(\omega-\mu_{\rm  L})}{2}\right)\cosh\left(\frac{\beta_{\rm R}(\omega-\mu_{\rm  R})}{2}\right)}\right],
        \\
    I_{\rm pair}&=2\int_{-\infty}^{\infty}
    \frac{d\omega}{2\pi}\sum_{\bm{q},\bm{q}'}
    |w_{\bm{q},\bm{q}'}|^2
    \mathcal{B}_{\bm{q},{\rm L}}(\omega)
    \mathcal{B}_{\bm{q}',{\rm R}}(\omega)
    \left[-\frac{1}{2}\frac{\sinh\left(\frac{\beta_{\rm b, L}(\omega-\mu_{\rm b, L})-\beta_{\rm R}(\omega-\mu_{\rm b, R})}{2}\right)}{\sinh\left(\frac{\beta_{\rm L}(\omega-\mu_{\rm  b,L})}{2}\right)\sinh\left(\frac{\beta_{\rm b,R}(\omega-\mu_{\rm b,R})}{2}\right)}\right],
    %b_{\rm L}(\omega),
\end{align}
\begin{align}
\label{eq:app_S2}
    \mathcal{S}_{\rm qp}&=\int_{-\infty}^{\infty}
    \frac{d\omega}{2\pi}\sum_{\bm{k},\bm{p},\sigma}
    |t_{\bm{k},\bm{p}}|^2\mathcal{A}_{\bm{k},{\rm L}}(\omega)
    \mathcal{A}_{\bm{p},{\rm R}}(\omega)
        \left[-\frac{1}{2}\frac{\cosh\left(\frac{\beta_{\rm L}(\omega-\mu_{\rm L})-\beta_{\rm R}(\omega-\mu_{\rm R})}{2}\right)}{\cosh\left(\frac{\beta_{\rm L}(\omega-\mu_{\rm  L})}{2}\right)\cosh\left(\frac{\beta_{\rm R}(\omega-\mu_{\rm  R})}{2}\right)}\right], \\
    \mathcal{S}_{\rm pair}&=4\int_{-\infty}^{\infty}
    \frac{d\omega}{2\pi}
    \sum_{\bm{q},\bm{q}'}|w_{\bm{q},\bm{q}'}|^2\mathcal{B}_{\bm{q},{\rm L}}(\omega)
    \mathcal{B}_{\bm{q}',{\rm R}}(\omega)
    \left[-\frac{1}{2}\frac{\cosh\left(\frac{\beta_{\rm L}(\omega-\mu_{\rm b,L})-\beta_{\rm R}(\omega-\mu_{\rm b,R})}{2}\right)}{\sinh\left(\frac{\beta_{\rm L}(\omega-\mu_{\rm b,L})}{2}\right)\sinh\left(\frac{\beta_{\rm R}(\omega-\mu_{\rm b,R})}{2}\right)}\right].
\end{align}
In particular, considering the large-biased limit where
\begin{align}
    \tanh\left(\frac{\beta_{\rm L}(\omega-\mu_{\rm L})-\beta_{\rm R}(\omega-\mu_{\rm R})}{2}\right)\simeq\tanh\left(\frac{\beta_{\rm L}(\omega-\mu_{\rm b,L})-\beta_{\rm R}(\omega-\mu_{\rm b,R})}{2}\right)\simeq 1,
\end{align}
is satisfied, we obtain 
\begin{align}
\mathcal{S}_{\rm qp}(\Delta\mu\rightarrow\infty)\rightarrow I_{\rm qp}, \qquad \mathcal{S}_{\rm pair}(\Delta\mu\rightarrow\infty)\rightarrow 2 I_{\rm pair},
\label{eq:fanofactorlimit}
\end{align}
where we have denoted $I\equiv I_{\rm qp}+I_{\rm pair}$.
The result of Eq.~(\ref{eq:fanofactorlimit}) motivates us to consider the Fano factor
\begin{align}
    F=\frac{\mathcal{S}}{I}=\frac{\mathcal{S}_{\rm qp}+\mathcal{S}_{\rm pair}}{I_{\rm qp}+I_{\rm pair}}.
\end{align}
Then, one can see that the Fano factor $F$ in a large-biased junction changes from 1 to 2 reflecting the ratio between $I_{\rm qp}$ and $I_{\rm pair}$.

\section*{Many-body $T$-matrix approximation}
To demonstrate this, we employ the many-body TMA to calculate spectral functions $\mathcal{A}_{\bm{k},{\rm j}}(\omega)$, $\mathcal{B}_{\bm{q},{\rm j}}(\omega)$, and $\mu_{\rm j}$ for given densities $N_{\rm j}$ in the BCS--BEC crossover regime~\cite{OHASHI2020103739}.
The single-particle propagator is given by 
\begin{align}
    G_{\bm{k},{\rm j}}(i\omega_n)&=\frac{1}{G_{\bm{k},{\rm j}}^0(i\omega_n)^{-1}-\Sigma_{\bm{k},{\rm j}}(i\omega_n)},\\
    \Sigma_{\bm{k},{\rm j}}(i\omega_n)
&=T_{\rm j}\sum_{\bm{q},\ell}\Gamma_{\bm{q},{\rm j}}(i\nu_\ell)G_{\bm{q}-\bm{k},{\rm j}}^0(i\nu_\ell-i\omega_n),
\end{align}
where $G_{\bm{k},{\rm j}}^0(i\omega_n) =(i\omega_n-\xi_{\bm{k},{\rm j}})^{-1}$ denotes the bare propagator and $\Sigma_{\bm{k},{\rm j}}(i\omega_n)$ denotes the TMA self-energy.
Following a standard TMA procedure~\cite{zwerger2011bcs}, the $T$-matrix $\Gamma_{\bm{q},{\rm j}}(i\nu_\ell)$ is formulated by incorporating the particle--particle multiple scattering as
\begin{align}
\label{eq:gamma}
\Gamma_{\bm{q},{\rm j}}(i\nu_\ell)
&=g\left[1-g\Pi_{\bm{q},{\rm j}}(i\nu_\ell)\right]^{-1},
\end{align}
using the bare two-body propagator given as
\begin{align}
    \Pi_{\bm{q},{\rm j}}(i\nu_\ell)
&=-T_{\rm j}\sum_{\bm{p},n}G_{\bm{p}+\bm{q}/2,{\rm j}}^0(i\omega_n+i\nu_\ell)G_{-\bm{p}+\bm{q}/2,{\rm j}}^0(-i\omega_n).
\end{align}
The fermion (boson) Matsubara frequency is denoted by $\omega_n$ ($\nu_\ell$).
Furthermore, we define the dressed two-body propagator~\cite{PhysRevB.61.15370} as
\begin{align}
    \mathcal{G}_{\bm{q},{\rm j}}(i\nu_\ell)=\Pi_{\bm{q},{\rm j}}(i\nu_\ell)\left[1+\Pi_{\bm{q},{\rm j}}(i\nu_\ell)\Gamma_{\bm{q},{\rm j}}(i\nu_\ell)\right].
\end{align}
The spectral functions can be obtained from the analytic continuation as $\mathcal{A}_{\bm{k},{\rm j}}(\omega)=-2\, {\rm Im}\, G_{\bm{k},{\rm j}}(i\omega_n\rightarrow\omega-\mu_{\rm j}+i\eta)$ and $\mathcal{B}_{\bm{q},{\rm j}}(\omega)=-2\, {\rm Im}\, \mathcal{G}_{\bm{q},{\rm j}}(i\nu_\ell\rightarrow \omega-\mu_{{\rm b},{\rm j}}+i\eta)$ with an infinitesimal small number $\eta$.

\section*{Retarded propagators in the dilute reservoir}

For the single-particle Green's function in the reservoir ${\rm R}$ at dilute limit, we employ the non-interacting one given by
\begin{align}
    G_{\bm{p},{\rm R}}^{\rm ret.}(\omega)=\frac{1}{\omega+i\eta-\epsilon_{\bm{p}}},
\end{align}
where the self-energy correction is ignored
[noting $\epsilon_{\bm{p}}=p^2/(2m)$].
For the two-body sector, we can rewrite the lowest-order two-body propagator as
\begin{align}
    \Pi_{\bm{q},{\rm j}}^{\rm ret.}(\omega)\equiv\Pi_{\bm{q},0}(\omega)
+\Xi_{\bm{q},{\rm j}}(\omega),
\end{align}
where
\begin{align}
    \Pi_{\bm{q},0}(\omega)
    &=\sum_{\bm{p}}\frac{1}{\omega+i\eta-\epsilon_{\bm{p}+\bm{q}/2}-\epsilon_{-\bm{p}+\bm{q}/2}}
\end{align}
and
\begin{align}
    \Xi_{\bm{q},{\rm j}}(\omega)
        &=-\sum_{\bm{p}}\frac{f_{{\rm j}}(\epsilon_{\bm{p}+\bm{q}/2})+f_{{\rm j}}(\epsilon_{-\bm{p}+\bm{q}/2})}{\omega+i\eta-\epsilon_{\bm{p}+\bm{q}/2}-\epsilon_{-\bm{p}+\bm{q}/2}}
\end{align}
are the in-vacuum two-body Green's function and the medium correction, respectively (for more details, see e.g., Refs.~\cite{STRINATI20181,OHASHI2020103739}).
Taking $\alpha^2=q^2/4-m\omega-i\delta$, we can analytically obtain
\begin{align}
      \Pi_{\bm{q},0}(\omega)  
    &=-\frac{m\Lambda}{2\pi^2}+\frac{m\alpha}{2\pi^2}\tan^{-1}\left(\frac{\Lambda}{\alpha}\right),
\end{align}
where $\Lambda$ is an ultraviolet cutoff.
Note that $\Lambda$ is renormalized via
\begin{align}
    \frac{m}{4\pi a}=\frac{1}{g}+\frac{m\Lambda}{2\pi^2},
\end{align}
which leads to
\begin{align}
\frac{1}{\Gamma_{\bm{q},{\rm j}}^{\rm ret.}(\omega)}&=\frac{m}{4\pi a}-\Pi_{\bm{q},{\rm j}}^{\rm ret.}(\omega)-\frac{m\Lambda}{2\pi^2}\cr
&\simeq\frac{m}{4\pi a}-\Xi_{\bm{q}}(\omega)-\frac{m\alpha}{4\pi}
\end{align}
where the ultraviolet divergence is cancelled ($\tan^{-1}\left(\frac{\Lambda}{\alpha}\right)\simeq \pi/2$ is used in the second line).
\par
In the dilute limit, the fermionic medium correction $\Xi_{\bm{q},{\rm R}}(\omega)$ is negligible. 
In this case, one can approximately obtain
\begin{align}
    \mathcal{G}_{\bm{q},{\rm R}}^{\rm ret.}(\omega)
    &\simeq\Pi_{\bm{q},0}(\omega)\left[1-g\Pi_{\bm{q},0}(\omega)\right]^{-1}.
\end{align}
where $\mathcal{G}_{\bm{q},{\rm R}}^{\rm ret.}(\omega)$ does not involve any poles on the real frequency axis (i.e. bound states) at $a^{-1}<0$.
Note that the two-body continuum exists above $\omega=q^2/(4m)$.
In the weak-coupling side ($a<0$), we obtain
\begin{align}
\mathcal{B}_{\bm{q},{\rm R}}(\omega)=-2\,{\rm Im}\,\mathcal{G}_{\bm{q},{\rm R}}^{\rm ret.}(\omega)=0. \quad (\omega <q^2/4m).
\end{align}
Simultaneously, the frequency integration is restricted as $\omega>0$.
This fact indicates that particles in the reservoir ${\rm L}$ are transferred to the two-body continuum in the reservoir ${\rm R}$ via the two-body tunneling process in the weak-coupling side ($a<0$).  
On the other hand, in the strong-coupling limit ($a\rightarrow +\infty$), we obtain~\cite{PhysRevB.61.15370,PhysRevB.68.144507}
\begin{align}
    \mathcal{G}_{\bm{q},{\rm R}}^{\rm ret.}(\omega)
    &\simeq \left(\frac{m\Lambda}{2\pi^2}\right)^2
        \frac{8\pi}{m^2a}\frac{1}{\omega+i\eta -\frac{q^2}{4m} + E_{\rm b}} \quad (\Lambda\rightarrow \infty),
\end{align}
which is proportional to the bosonic Green's function of a bound molecule with the binding energy $E_{\rm b}=1/(ma^2)$.
Thus, in the strong-coupling regime ($a>0$),
particles in the reservoir ${\rm L}$ can be transferred to the molecular bound states in the reservoir ${\rm R}$ via the two-body tunneling process.

\section*{Large-bias limit}

In the main text, we considered a situation where fermions in the strongly-correlated reservoir ${\rm L}$ with a finite density $N_{\rm L}$ go through the tunneling junction to the dilute reservoir ${\rm R}$ with a vanishing density $N_{\rm R}\rightarrow 0$, i.e., $\mu_{\rm R}\rightarrow -\infty$. While we take the same temperatures $T_{\rm L}=T_{\rm R}$ in the two reservoirs, $T_{\rm R}$ does not affect the result in the present case of $\mu_{\rm R}\rightarrow -\infty$ because the fugacity $z_{\rm R}=e^{\mu_{\rm R}/T_{\rm R}}$ characterizing the distribution vanishes regardless of the value of $T_{\rm R}$.
The condition of the large bias limit [$\mu_{\rm R}=T_{\rm R}\ln(z_{\rm R})\rightarrow -\infty$] is unchanged in both BCS and BEC sides at nonzero temperatures because the dilute reservoir ${\rm R}$ obeys the Boltzmann statistics.
Indeed, we obtain vanishing $N_{\rm R}$ as~\cite{PhysRevA.91.013606}
\begin{align}
 N_{\rm R}= 2z_{\rm R}
 \left(\frac{2\pi}{mT_{\rm R}}\right)^{\frac{3}{2}}
 +O(z_{\rm R}^2)\rightarrow 0 \quad(z_{\rm R}\rightarrow 0).   
\end{align}
The number density $N_{\rm L}$ of the ${\rm L}$-reservoir can be numerically obtained from
\begin{align}
    N_{\rm L}=T_{\rm L}\sum_{\bm{p},\sigma,n}G_{\bm{p},{\rm L}}(i\omega_n).
\end{align}
In this regard, we normalize physical quantities by using the Fermi energy $E_{\rm F,L}=(3\pi^2 N_{\rm L})^{\frac{2}{3}}/(2m)$ and momentum $k_{\rm F,L}=(3\pi^2 N_{\rm L})^{\frac{1}{3}}$.

%\begin{figure}[t]
%    \centering
%    \includegraphics[width=7.5cm]{fig_mudep.eps}
%    \caption{\red{The inverse Fano factor $F^{-1}=I/\mathcal{S}$ as a function of the chemical-potential bias $\Delta\mu=\mu_{\rm L}-\mu_{\rm R}$ at $T_{\rm L}/T_{\rm F,L}=0.3$, $a^{-1}=0$, and $\mathcal{T}_{2,{\rm ren.}}/\mathcal{T}_1=1$.
%    The dashed and dotted lines represent Onsager's relation given by Eq.~(\ref{eq:Onsager}) and the large-bias limit ($\mu_{\rm R}\rightarrow-\infty$), respectively.}}
%    \label{fig:mudep}
%\end{figure}

%\red{
%Figure~\ref{fig:mudep} shows $\Delta\mu$-dependence of the inverse Fano factor $F^{-1}=I/\mathcal{S}$, where the parameters are chosen as $T_{\rm L}/T_{\rm F,L}=0.3$, $a^{-1}=0$, and $\mathcal{T}_{2,{\rm ren.}}/\mathcal{T}_1=1$.
While the Fano factor is well described by the Onsager's relation $F^{-1}(\Delta\mu\rightarrow 0)=\frac{\Delta\mu}{2T}$ in the low-bias regime, $F^{-1}$ approaches the large-bias limit ($\mu_{\rm R}\rightarrow -\infty$) when $\Delta\mu/ E_{\rm F,L}\gesim 1$.
This indicates that it is sufficient to reach the large-bias limit when $\Delta\mu$ is larger than the many-body scale of the reservoir, that is, $E_{\rm F,L}$. 
We note that $\Delta\mu$ can be controllable in cold atomic experiments by preparing the reservoirs with the large density imbalance.

\section{Different tunneling-coupling ratio}

\begin{figure}[t]
    \centering
    \includegraphics[width=7.5cm]{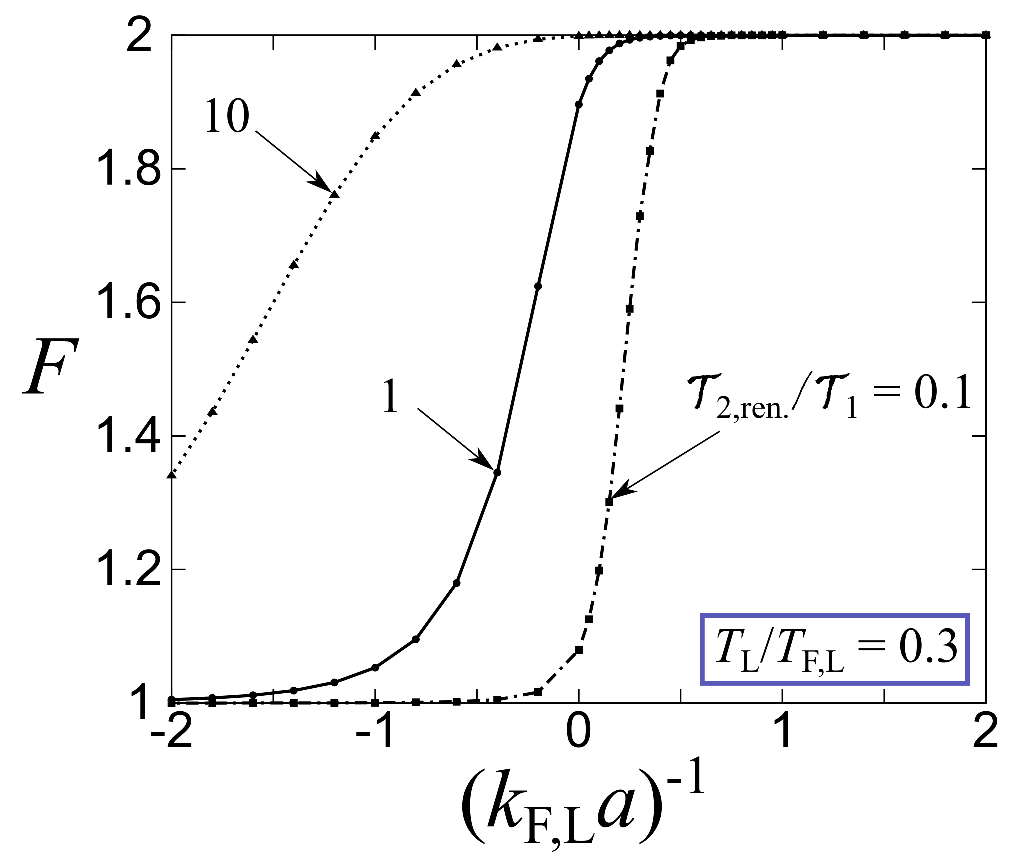}
    \caption{Fano factor $F$ throughout the BCS-BEC crossover at different tunneling-coupling ratio $\mathcal{T}_{2,{\rm ren.}}/\mathcal{T}_1$. The temperature is taken as $T_{\rm L}/T_{\rm F,L}=0.3$.
    One can see that $F$ changes from 1 to 2 with increasing the interaction strength regardless of different $\mathcal{T}_{2,{\rm ren.}}/\mathcal{T}_1$. }
    \label{fig:tratio}
\end{figure}

Figure~\ref{fig:tratio} shows the calculated Fano factor $F$ with different 
tunneling-coupling ratio $\mathcal{T}_{2,{\rm ren.}}/\mathcal{T}_1$ in the entire BCS-BEC crossover regime at $T_{\rm L}/T_{\rm F,L}=0.3$.
While in the main text we employed $\mathcal{T}_{2,{\rm ren.}}/\mathcal{T}_1=1$,
this ratio depends on the actual detailed setups in each experiment.
If the two-body tunneling is relatively strong as 
$\mathcal{T}_{2,{\rm ren.}}/\mathcal{T}_1=10$, $F$ is close to 2 even in the weak-coupling side [$(k_{\rm F,L}a)^{-1}\simeq -1$].
However, $F$ decreases at weaker coupling even in this case.
On the other hand, in the case with $\mathcal{T}_{2,{\rm ren.}}/\mathcal{T}_1=0.1$,
$F$ remains to be close to 1 even around unitarity.
Nevertheless, $F$ rapidly increases around $(k_{\rm F,L}a)^{-1}=0.3$ and consequently reaches $F=2$ in the strong-coupling limit.

In this way, the detailed structure of the tunneling junction affects how $F$ increases in the BCS-BEC crossover regime.
However, our conclusion that $F=1$ and $F=2$ are achieved in the BCS and BEC limits, respectively, is unchanged even for different tunneling-coupling ratios.
In other words, the pair tunneling process inevitably occurs in the strong-coupling regime even for an infinitesimally small pair-tunneling coupling $\mathcal{T}_{2}$.
This is a natural consequence in the sense that the system is dominated by bound molecules and hence there are no single-particle states in such a regime.

We note that the value of $\mathcal{T}_{2,{\rm ren.}}/\mathcal{T}_1$ is associated with the potential barrier and the interaction strength~\cite{PhysRevA.106.033310}. While it is not so straightforward to estimate $\mathcal{T}_{2,{\rm ren.}}/\mathcal{T}_1$ in each experimental setup,
it is sufficient to observe $F$ at the regime where the anomalously large tunneling current can be found [e.g., at unitarity observed in Ref.~\cite{PhysRevLett.126.055301}] for our purpose of detecting the pair-tunneling current.

\end{widetext}

\end{document}